\documentclass[10pt,letterpaper,table]{article}
\usepackage[top=0.85in,left=2.75in,footskip=0.75in]{geometry}

\usepackage{amsmath,amssymb}

\usepackage{changepage}

\usepackage{textcomp,marvosym}

\usepackage{cite}

\usepackage{nameref}
\usepackage[hidelinks]{hyperref}
\usepackage{orcidlink}
\usepackage{rorlink}

\usepackage[nopatch=eqnum]{microtype}
\DisableLigatures[f]{encoding = *, family = * }

\usepackage{array}

\newcolumntype{+}{!{\vrule width 2pt}}

\newlength\savedwidth%

\raggedright%
\usepackage[aboveskip=1pt,labelfont=bf,labelsep=period,justification=raggedright,singlelinecheck=off]{caption}

\makeatletter
\renewcommand{\@biblabel}[1]{\quad#1.}
\makeatother

\usepackage{lastpage,fancyhdr,graphicx}
\usepackage{epstopdf}
\fancyheadoffset[L]{2.25in}
\fancyfootoffset[L]{2.25in}
\usepackage[inline]{enumitem}

\begin{document}
\vspace*{0.2in}

\begin{flushleft}
{\Large
\textbf\newline{Ten simple rules for PIs to integrate Research Software Engineering into their research group} 
}
\newline
\\
Stuart M. Allen\orcidlink{0000-0003-1776-7489}\textsuperscript{1\Yinyang},
Neil Chue Hong\orcidlink{0000-0002-8876-7606}\textsuperscript{2,3\Yinyang},
Stephan Druskat\orcidlink{0000-0003-4925-7248}\textsuperscript{4\Yinyang},
Toby Hodges\orcidlink{0000-0003-1766-456X}\textsuperscript{5\Yinyang*},
Daniel S. Katz\orcidlink{0000-0001-5934-7525}\textsuperscript{6\Yinyang},
Jan Linxweiler\orcidlink{0000-0002-2755-5087}\textsuperscript{7,8\Yinyang},
Frank Löffler\orcidlink{0000-0001-6643-6323}\textsuperscript{8,9\Yinyang},
Lars Grunske\orcidlink{0000-0002-8747-3745}\textsuperscript{10\Yinyang},
Heidi Seibold\orcidlink{0000-0002-8960-9642}\textsuperscript{11\Yinyang},
Jan Philipp Thiele\orcidlink{0000-0002-8901-6660}\textsuperscript{8,12\Yinyang},
Samantha Wittke\orcidlink{0000-0002-9625-7235}\textsuperscript{13\Yinyang}
\\
\bigskip
\textbf{1} School of Computer Science and Informatics, Cardiff University, Cardiff, Wales, United Kingdom\rorlink{https://ror.org/03kk7td41}
\\
\textbf{2} EPCC, University of Edinburgh, Edinburgh, Scotland, United Kingdom\rorlink{https://ror.org/01nrxwf90}
\\
\textbf{3} Software Sustainability Institute, United Kingdom\rorlink{https://ror.org/0455awd29}
\\
\textbf{4} German Aerospace Center (DLR), Berlin, Germany\rorlink{https://ror.org/04bwf3e34}
\\
\textbf{5} The Carpentries, Oakland, California, USA\rorlink{https://ror.org/0356fgm10}
\\
\textbf{6} NCSA \& Siebel School of Computing and Data Science \& School of Information Sciences, University of Illinois Urbana-Champaign, Urbana, IL, USA\rorlink{https://ror.org/047426m28}
\\
\textbf{7} Technische Universität Braunschweig, Germany\rorlink{https://ror.org/010nsgg66}
\\
\textbf{8} de-RSE e.V. - Society for Research Software, Germany\rorlink{https://ror.org/007qpef44}
\\
\textbf{9} Competence Center Digital Research, Friedrich Schiller University Jena, Germany\rorlink{https://ror.org/05qpz1x62}
\\
\textbf{10} Department of Computer Science, Humboldt-Universität zu Berlin, Germany\rorlink{https://ror.org/01hcx6992}
\\
\textbf{11} Digital Research Academy, Munich, Germany
\\
\textbf{12} Weierstrass Institute Berlin, Germany\rorlink{https://ror.org/00h1x4t21}
\\
\textbf{13} CSC- IT Center for Science, Espoo, Finland\rorlink{https://ror.org/04m8m1253}
\\
\bigskip

%
%
\Yinyang\ These authors contributed equally to this work as first author.





*\ tobyhodges@carpentries.org

\end{flushleft}





\section{Introduction}
Research Software Engineering (\textit{RSEng}) is a key success factor in producing high-quality research software~\cite{cohen_etal_2021_four_pillars_research} 
and thus improves research project outcomes, since better research software leads to better research~\cite{gobleBetterSoftwareBetter2014}. 
However, as a leader of a research group or project you may not know what RSEng is, or how you can maximize its impact on your research.  
To make matters worse, if you try to read about RSEng or strike up a conversation about how it might be relevant to your research, you might be met with complicated technical details~\cite{SWEBOK}. 
Surely, it must be possible to learn what RSEng is about without first enrolling in a  Computer Science program at your university!

Many explanations and tutorials for development aspects like testing, software architecture and version control are very technical. 
This prevents researchers and other decision makers from making an informed choice about employing these methods. 
The result may be research software that is less robust~\cite{taschuk_wilson_2017_ten_simple_rules}
and less usable~\cite{hunter-zinck_etal_2021_ten_simple_rules}, 
and research that is less reproducible~\cite{sandve_etal_2013_ten_simple_rules}.
In contrast, using fundamental RSEng methods can make code good enough,
even when programming needs to be quick and dirty~\cite{balaban_etal_2021_ten_simple_rules}.
Our aim is to provide comprehensible descriptions of simple rules to improve software-enhanced research.




\section{The Rules}

It is important to apply appropriate Software Engineering concepts and use applicable Software Engineering practices~\cite{SWEBOK} to the development of research software. As a leader of a research project or group, you should check that your team follows these 10 RSEng rules in order to produce amazing, reproducible, and trustworthy research software and results.

\subsection*{Rule 1: Use Research Software Engineering methods and processes from the start} 

A common thread linking many of our rules is that their adoption is part of a process of development and not a sort of `clean-up phase' at the end of a project. 

Defining the requirements of the software (Rule 3) and planning its architecture (Rule 4) at the start of the project, then revisiting those aspects throughout, ensures that software engineering efforts are directed and remain aligned with the goals of your research. Considering the specific needs of the research project and how the individual parts of the software should fit together only at a later time might lead to a catastrophic misalignment.

Spending time early on to identify the appropriate tools and methods (Rule 5) to use in the software reduces the likelihood of progress stalling later in the project when unanticipated limitations are encountered.

Version control (Rule 6) and documentation (Rule 8) for specific research software can be seen as the lab notebook of the computational scientist. 
As an experimental scientist takes notes throughout a research project, computational researchers need time to take the corresponding notes for their software projects as they are developed. Drafting them at a later date, when the developer is less in touch with their thought process (or those of their collaborators), is more difficult, time-consuming, and prone to mistakes and omissions.

Developing the software in the open (Rule 7) from the very beginning maximises the time in which potential collaborators can find the project and contribute to its development. Taking this step only after the software has already been developed is less likely to result in fruitful collaborations.

Tests that are designed and adjusted as the research software is developed (Rule 9) ensure its ongoing quality and accuracy, saving time and resources that may otherwise be lost when an erroneous result goes undetected.

Providing citation information (Rule 10) early and updating it often enables users to credit your research group whenever they use the software. Collating this information after the fact is usually slower and more difficult than collecting it as the software grows.

As research software experts, Research Software Engineers (\textit{RSEs}) can apply suitable methods at the right time in the development process and lead by example.
This will result in software that is maintainable, manageable, reusable, and adaptable, saving resources in the long run.

\subsection*{Rule 2: Research Software Engineering is (much) more than programming (writing code)}
RSEs have previously been stereotyped as lone researchers, with no software engineering training,
``hacking'' late into the night to produce code that gives results at the cost of other factors such as robustness or maintainability. 
However, the process of developing research software properly involves much more than just the activity of writing lines of code to implement the desired functionality. 
This is similar to software engineering in general,
which includes quality assurance, maintenance, management, operations, and economics.

The activities involved in RSEng can be considered in three categories:
\begin{itemize}
    \item Technical activities: e.g., writing code, architecture and design, testing, optimisation, porting to new platforms;
    \item Management activities: e.g., team management, project management, product management;
    \item Knowledge transfer activities: e.g., documentation, training, community building, mentoring, user support.
\end{itemize}

In each of these categories, the activities must be carried out with an understanding of the specific research needs (see Rule 3). This can place a greater emphasis on factors such as reproducibility, collaboration, and supporting funding applications or conference/publication submission deadlines.

RSEng work can be performed by dedicated RSEs or by researchers who code and ideally follow software engineering principles. In both cases, research leaders and institutions should recognise the importance of RSEs as a key component of successful and sustainable research, providing opportunities for career progression, training, and support.

\subsection*{Rule 3: Translate the needs of the research into the needs of the software}
\label{subsec:researchneeds}

Successful enterprise software projects are founded on the a priori elicitation, modelling, and management of stakeholder requirements, 
allowing developers to deliver solutions that meet the defined needs of their end users~\cite{Hofmann2001}.  
In contrast, research software is often developed either without clear requirement specifications or with only a limited focus on high-level requirements. 
Developers are left to prioritise between these, often with limited and infrequent input from the end user for whom the software is being built~\cite{Heaton2015}. 
As a result~\cite{bajraktari_etal_2024_requirements_engineering_researcha}, research software can fail to satisfy the needs of all its potential stakeholders (including other researchers, reviewers, funders, and wider society), 
potentially leading to software that is not reusable and/or interoperable. 

Best practices from industry do not always translate to research software development, 
for example, due to small, often non-dedicated teams with a lack of formal software engineering training.
Nevertheless, flexible and dynamic methodologies such as Agile Development~\cite{agile} can and should be implemented and adapted to this context. The goals of the underlying research should be integrated into the software development process.



\subsection*{Rule 4: Think about the software architecture before and while you write code}
\label{subsec:architecture}

The development of research software is often 
exploratory, for example when
a research group has a hypothesis to test, an experiment to run, or a paper to write. 
Code may be 
developed without a clear idea or discussion 
of how the requirements for the software
are mirrored in its design. 
At the end of such a process, the software will have an architecture,
but it will be accidental and undocumented.
This makes it harder to build and use the software and to change it later.
It will be unclear, for example,
how different software parts relate and interact with each other,
and what effect any changes made in one part will have on other parts.

By explicitly designing and documenting the software architecture,  
decisions can be
discussed and agreed upon before writing and building the software,
and before using it.

The architecture of a piece of software 
is the conceptual glue that
helps communicate what is required 
to maintain
the goals of the software and achieve new ones as it evolves
and as stakeholders change.
Much like how physical experiments are performed by combining multiple devices, such as sensors, into a complete setup,
computational (in silico) experiments are performed by combining software components into an executable application.
The software architecture that says how this combination happens is similar to the architectural plans for a house:
a way for the people who will live in the house to visualise it, and identify potential issues; 
a design that makes it easier for the builders to construct it;
and a plan that makes it easier to repair or extend in the future.

Software architecture makes research software more maintainable, extensible, and secure~\cite{druskat2025architecture}.
This is especially important if the software is used for multiple research projects, by different collaborators, or by the wider research community. 
A documented software architecture makes it easier for new members to join the project: together with documentation and tests, it helps understand how the software is constructed, what it does, and where and how additional contributions can be integrated.

There is no single right way to document research software architecture.
You can start with a lightweight, high-level description of the essentials, or with a diagram.
It is more important that you consider and discuss architecture early in the development process,
and that your group collaborates to review, evolve, and document the architecture together with your code.


\subsection*{Rule 5: Use the right tool for the job}

It is common to hear software developers speak of ``the right tool for the job'', often in negative statements, for example, when something ``is not the right tool for the job''. 
The assumption that a single ``right tool'' exists is usually wrong. 
In practice, the list of candidate tools for any software task may be long, 
and the advantages and disadvantages of each must be evaluated in the context of the research project, the current expertise of your research group, and the established norms of the research domain.

When planning the development of research software, the project lead, in consultation with colleagues and collaborators in the research project, must consider the following:
\begin{enumerate*}
 \item How well does the job fit the typical or intended usage of the tool? 
    Those do not necessarily need to overlap, but it is a good idea to be more careful when they do not overlap.
 \item What factors might exclude some choices? 
    Relevant considerations include cost, license incompatibilities, availability in the target environment, desired user interface, and performance.
 \item How familiar are your team members with a given tool? 
    Inflexibility can be dangerous here, leading to seeing everything as a nail when you have a hammer, but it also not to be disregarded lightly. 
    Time and work required to learn new tools and approaches can be well invested, but must be balanced against the other needs of the research project. 
    An unfamiliar tool might be best-suited to a given job, 
    but if your team needs a lot of time to learn it and an alternative they already know is almost as good, the ``right'' tool for the job is not the ``best''.
 \item How easy would it be for your team members to get help when you face challenges with a given tool? 
    At some point, everyone needs help with any tool. 
    The quicker and easier they can get help, the better they can do their job, and the more likely this tool is ``right''.
 \item Who else is affected by the choice and what are their answers to the questions above? 
    All of the above considers only your research team, 
    but research software is much more likely to be adopted if it fits easily into the ecosystem of other tools that is already established in your domain.
\end{enumerate*}

The bottom line is that there is a lot to consider and much of that is not about the job itself, but its context.

\subsection*{Rule 6: Version control tracks changes and averts disasters}

Software is usually developed incrementally, with its functionality, behaviour, and design changing and improving as new features and better methods are introduced, bugs are found and fixed, and code is improved. 
This is certainly true for research software, where our needs from software can grow and change with the findings and methodology of the research project at large.

Like ``track changes'' for software, version control---usually enabled by the program git and the ecosystem of tools that has evolved around it---empowers developers to record and document changes made to software by multiple developers over time.
It also makes it much easier to compare software versions, and to find, then discard or revert changes if and when developers realise that they have introduced mistakes.
Version control tools allow fast and frequent synchronisation of project files across systems, effectively supporting distributed collaboration and providing additional distributed backups while avoiding the need to email files back and forth.

Version control saves researchers huge amounts of time by facilitating experimentation, especially when adopted from the very beginning of the development process. 
Version control becomes even more helpful when the research software development process is collaborative.
Git (\url{https://git-scm.com/}], and popular platforms that host software projects using it, such as GitLab (\url{https://gitlab.com/}) and GitHub (\url{https://github.com/}), 
facilitate collaboration on code and software projects in much the same way as tools like Online Office Suites, web-based LaTeX editors, or Markdown Pads have made it easier for multiple authors to write documents together.
Without a version control system, you often end up with developers having various versions of the same software that diverge from each other.
\textbackslash{}emph\{Continuous Integration\} techniques can help to manage frequent contributions from multiple authors, for example, by running automated test cases as revised code is submitted to avoid conflicts.

\subsection*{Rule 7: Open source facilitates collaboration and open science}

Software is \emph{open-source} when its source code is publicly available under a license that meets the requirements of the Open Source Initiative~\cite{OSD} or the Free Software Foundation~\cite{FSF}, including that it allows reuse and modification by others. 

Using open-source software has several benefits.
First, as open-source software is published under a license designed to support reuse, it can be used freely.
A lot of open-source software projects are sustained by a community of contributors, which can mean that documentation and support are easily available online.
Devoting time of your group to contribute to the development and maintenance of the software you use in your research is a good way to ensure its ongoing availability, 
to help researchers develop their own software engineering skills, and to establish collaborations.
Using open-source software, licensed to encourage reuse, greatly facilitates its extension and modification for new and interesting research directions you might not think of yourself.
As such, open-source is one important part of open science.
Finally, access to the software used in research is necessary to reproduce that research.

Developing research software openly under an open-source license makes it easier for others to reuse your work.
Increased adoption of your work leads to more impact for your research, more collaboration, and more citation~\cite{mckiernan2016open}.
Additionally, it can pay dividends to ``plug into'' an existing open source tool/set of tools when creating research software.
Reusing existing software avoids ``reinventing the wheel,'' allowing you to focus your efforts on the novel methods that your research demands, 
as a single research team typically can't cover the full depth and breadth of methods and algorithms needed. 

Using platforms such as GitHub and GitLab facilitates collaboration within and between teams, especially across multiple time zones.
These platforms remove barriers to ongoing contributions from anyone, anywhere, world-wide.
They also support researchers to continue working with their software when they change institutions~\cite{littauer202510quicktipsmaking}.
Finally, maintaining your research software as an open source project and building a community of developer around it can increase its longevity.

Your publications also benefit from your software being open source.
Providing the source code alongside your manuscript, as well as the associated data and other artifacts of your research, helps journal editors, reviewers, and everyone else understand your methodology, reproduce your results, and build on your research.

\subsection*{Rule 8: Documentation is essential to research software}%
\label{sec:documentation}

Documentation is material that accompanies software, describing what it does, how it is structured, how to use it, and potentially how to contribute to it. 
Documentation can exist in different forms for different audiences, e.g., tutorials for new users, quick reference guides for the more experienced, etc.

Without documentation, researchers and RSEs may not make use of existing research software, no matter how good the actual software is.
Research software, like any other software, needs good documentation in order to make it useful for the intended audience, meaning the documentation needs to be written with the audience in mind.
Developers (even those who wrote the software) need documentation to help them quickly understand or remember how the software is designed, implemented, and works.
Also, users, who may be researchers without software development experience, need to know how to use the software in their work.
RSEs know about the different types of documentation (e.g., as defined by the Diátaxis framework (\url{https://diataxis.fr/}) and how to formulate them towards different audiences (e.g., a user, a collaborating developer).
RSEs typically have experience within the domain that the research software is addressing and can provide a tutorial/examples that make sense and are useful to the domain researcher audience.
In some cases, a README is sufficient to document research software, but not typically.
RSEs know when to go from README to something more and how to make documentation approachable, accessible and [usable] (e.g., copy-pastable code blocks).

As research software often grows over time, documentation should be kept close to the code (and even embedded in it) and should be an essential part of the development process, not an afterthought.
This requires planning time to document the research software during the project planning phase.

The existence and quality of documentation can be the difference between researchers deciding to use or contribute to software.
Good documentation can save users and developers valuable time, which can otherwise be spent researching, writing grant proposals or extending the software instead of being ``helpdesk'' or digging into the code to find out how to use it.

\subsection*{Rule 9: Apply rigorous quality assurance techniques}

The quality of the research software directly influences the quality and trustworthiness of the research outputs that use it. 
Thus, rigorous software quality assurance (QA) practices need to be applied to ensure the reliability and validity of the research results. 
However, we know~\cite{CarverKSP07,KanewalaB14,VogelDSDG19} that pure testing of the scientific software is difficult, 
since we often do not know the correct output beforehand, e.g., in a novel experiment.
As a result, QA practices should be more fine-grained 
and ensure that research software behaves as expected 
and that its outputs are reliable across different environments. 
To be effective, QA practices should be part of the continuous integration/ continuous delivery (CI/CD) environment 
and should automatically run after each change added to the version control system 
or at least regularly, e.g., every night/week, and before each release of a new version.

The question now is, what quality assurance practices are effective and should be included? 
We believe that research software can benefit from all typical (automated) software quality assurance techniques, 
which includes both static and dynamic code analysis.
Static Analysis checks for the absence of known fault patterns, 
as well as adherence to the desired code style, without needing to run the code.
On the other hand, dynamic analysis checks for expected runtime behaviour, e.g., by running test cases for known scenarios with known results.
These tests should at least be run for each meaningful evolution step of the software.

The tests should cover a wide range of interactions within the software.
Since research software should be decomposed into manageable modules, 
as described in a previous rule (Rule 4),
most of the test should also be done at the level of these modules as so-called unit tests.
As the goal of modules is to have a manageable piece of software, 
writing test cases should be manageable as well.
Ideally, the implementation should follow the test-driven development idea, 
where developers specify the required/expected output for fixed/known module inputs.
Since research software is often built with interconnected modules as well, 
integration testing should be performed to check if the modules work together smoothly. 

The benefit of a CI/CD pipeline, which automatically reruns these unit and integration test cases, is
that future code changes do not break the intended behaviour of the module and the entire research software unnoticed.

Encouraging RSEs to engage in frequent and constructive code review, where one RSE reads the software written by another and provides feedback, 
is another way to safeguard the quality of research software. 
When done well, code review improves the quality and consistency of software, fosters collaboration, and provides frequent opportunities for RSEs to learn new techniques and best practices. 
If your research group contains only one RSE, they might achieve this by connecting with colleagues in another group or a central RSE team at your institution.

Quality assurance is good for research in your group because:
    \begin{enumerate*}
        \item It saves time and money finding and solving problems;
        \item It increases confidence that the results your software produces are correct, reducing the likelihood of embarrassing retractions in your future and, oops, now I have made this a scary story you might tell PIs around a campfire;
        \item You would not use uncalibrated lab equipment, so why use untested software?
    \end{enumerate*}

\subsection*{Rule 10: Research software can and should be published and cited}

Research software created in an scholarly context should be subject to the same cultural and ethical practices 
as any other research activity and outcome: transparency, availability, traceability, etc. Some of these 
requirements are summarised in the FAIR (findable, accessible, interoperable and reusable) Principles, 
particularly the version adapted for research software~\cite{FAIR4RSprinciples}. To satisfy these principles, 
research software must be published in a suitable venue, with metadata that helps researchers find and access
the software itself, assess its interoperability, and evaluate its reusability.
The research software also needs a suitable license to allow others to reuse it and build on it.
RSEs and other software licensing experts can help you make 
an informed license choice, taking into account institutional policies and requirements.

Publishing your research software also creates an object in the scholarly record
that users can cite.
Citing research software is important because it gives credit to the people who have created the software.
Citing software helps identify which software has been used to conduct the research and produce the results, which contributes to making research reproducible.
Moreover, not citing the software that you use makes your research paper's methods section nearly useless.

Software publication can be similar to paper publication: once you want to release a version of your software, 
you should give it a version identifier and describe it with relevant metadata.
Then, you should deposit the snapshot of software artifacts (source code and/or executable files, documentation) 
in an archival repository such as Zenodo (\url{https://zenodo.org}), where it gets a unique identifier (a DOI).
To enable easier reuse, you can publish its required environment as well, e.g., 
in a virtual machine snapshot or a container that includes relevant parts of the operating system 
and other software packages that users need to run the software.

When citing software you use~\cite{SoftCitePrin}, be sure to include:
\begin{enumerate*}
    \item The correct list of authors, which may include people that haven’t written and committed a single line of code: designers, architects, testers, community builders, etc. (If the project hasn't provided this, it's better to use the project's name as the author rather than trying to determine the individual authors.)
    \item The name of the software.
    \item A means of identification of the specific version: ideally a persistent identifier but at least a combination of a version identifier and a release/publication date.
\end{enumerate*}

\section{Conclusion}
Some of these 10 simple rules highlight practical aspects designing and implementing high quality and sustainable research software; it's important  that the members of your research group developing software have the practical knowledge and skills to implement them, or the opportunities to gain those skills and knowledge. 
The other rules should be considered `good scientific practice'; everyone in your group should know them. 

As a final note there are many national and regional RSE societies (\url{https://researchsoftware.org/assoc.html}), 
which promote the discussion and sharing of knowledge about RSEng, e.g., at regular conferences. 
Engagement and participation from your research group is encouraged.

\section*{Acknowledgments}
We would like the thank Schloss Dagstuhl, Leibniz Center for Informatics, 
and everybody who attended the Dagstuhl Seminar ``24161 Research Software Engineering: Bridging Knowledge Gaps''.
Additionally we like to thank everyone who filled out the surveys on RSEng which led to some of the content of our rules.


\bibliography{references}

%
%
%
%
%
%
%

\end{document}